\documentclass{aastex}
\usepackage{spr-astr-addons}
\usepackage{url}

\RequirePackage{color}

\newcommand{\eqb}{\begin{equation}}
\newcommand{\eqe}{\end{equation}}
\begin{document}
\title{Adjustment of the electric current in pulsar magnetospheres and origin of subpulse modulation}

\shorttitle{Subpulse modulation}
\shortauthors{Short Author}

\author{Yuri Lyubarsky}
 \affil{Physics Department, Ben-Gurion University, P.O.B. 653, Beer-Sheva 84105, Israel}
 %\date{\today}

\begin{abstract}
The subpulse modulation of pulsar radio emission goes to prove that the plasma flow in the open field line tube breaks into isolated narrow streams. I propose a model which attributes formation of streams to the process of the electric current adjustment in the magnetosphere. A mismatch between the magnetospheric current distribution and the current injected by the polar cap accelerator gives rise to reverse plasma flows in the magnetosphere. The reverse flow shields the electric field in the polar gap and thus shuts up the plasma production process. I assume that a circulating system of streams is formed such that the upward streams are produced in narrow gaps separated by downward streams. The electric drift is small in this model because the potential drop in narrow gaps is small. The gaps have to drift because by the time a downward stream reaches the star surface and shields the electric field, the corresponding gap has to shift. The transverse size of the streams is determined by the condition that the potential drop in the gaps is sufficient for the pair production. This yields the radius of the stream roughly 10\% of the polar cap radius, which makes it possible to fit in the observed morphological features such as the "carousel" with 10-20 subbeams and the system of the core - two nested cone beams.
\end{abstract}


\maketitle
\section{Introduction}
The nature of pulsar emission is still a subject of debate.
The conventional wisdom is that the pulsar activity is associated with
relativistic plasma flows along the open magnetic field lines. According to the commonly accepted picture,
the polar cap cascade produces electron-positron
pairs (see, e.g., reviews by \citet{grenier_harding06, arons07} and
references therein), which stream along the open magnetic field lines with relativistic velocities. The pulsar radio emission is believed to be generated in this outflow.

Because of strong relativistic beaming, the basic morphological features of the pulsar radiation are dictated by the geometry of flow. It is generally accepted that the width of the mean pulse is determined by the width of the open field line tube in the emission region so that narrow pulses imply low emission altitude. The subpulse structure is commonly attributed to a systems of narrow streams of plasma, each stream radiating a subbeam tangentially to the magnetic field lines. These subbeams exhibit pretty regular behavior. It was found that pulsars have attractors in the form of two nested cones \citep{rankin83,rankin93}. Within the cone, the subpulse behavior also exhibits substantial regularity  such as stationary drifting \citep{rankin86,weltevrede06,weltevrese07}. In a couple of cases, a careful analysis of the data reveals a remarkable circular "carousel" of emitting subbeams \citep{deshpande_rankin01,mitra_rankin08}. We are still lacking of a firm model unequivocally explaining why the flow in the open field line tube breaks into isolated streams.

According to \citet{ruderman_sutherland75}, the electric field is unable to extract ions from the surface of the neutron star therefore electron-positron pairs are produced in local gap discharges (sparks).  Each spark gives rise to a narrow plasma stream responsible for a subpulse. The sparks are isolated one from another because the potential drop in the space between them remains below that necessary for the plasma production. This model remains the most popular ( see, e.g., recent paper by \citet{vanLeewen_timokhin12}). The basic problem with this model is that the surface binding energy was found to be not high enough to prevent a free supply of charges from the surface. Then the primary beam is slowly accelerating so that the gap height exceeds the radius of the polar cap \citep{muslimov_tsygan92,hibsch_arons01a,hibsch_arons01b}; in this case the electric separation of sparks does not work. \citet{gil_melikidze02,gil_melikidze_zhang06} propose to resolve this difficulty assuming that the surface magnetic field is significantly larger than the dipole.
%attempted to avoid this difficulty assuming that due to high multipole structure, the magnetic field at the surface is significantly higher than it has been estimated from the spin down rate.
%Moreover, this model attributes the subpulse drift to the $\mathbf{E\times B}$ drift, which cold hardly be reconciled with the observed drift reversals (e.g., \citet{nowakowski91,esamdin05}).

%The physics of the polar cap cascade is widely discussed .
\citet{wright03} proposed an empirical model for subpulse modulation based on the assumption that in the open field line tube, particles stream in both directions between the polar cap and the outer magnetosphere and these flows are separated one from another forming a slowly circulating system of plasma columns. With a number of other assumptions, this model nicely describes the observed subpulse behavior however, the model is still lacking of a solid physical background. Wright suggested that the reverse flows are produced by the pair cascades in outer gaps. It is not at all evident that outer gaps exist and could produce pairs, especially in slowly rotating pulsars, and
%If the polar cap cascade produces sufficient amount of plasma, no other gap arises in the outer magnetosphere.
in any case,  it is not clear why the plasma flows from the outer and polar gaps should be separated.
However, the principal idea that the formation of the subpulse system
%involves the entire magnetosphere  and
should be attributed to an intercommunication between the outer magnetosphere and the polar cap sounds reasonable.
 %statement that "events in the outer magnetosphere are the true "engine" of the pulsar system and that the observed polar cap region events are passive reflectors of these" seems to be.
An important point is that one need not look for an active region in the outer magnetosphere in order to find a feedback mechanism. Generally the structure of the magnetosphere is not driven by plasma producing gaps, neither by the polar nor by the outer ones, but on the contrary, the global electrodynamics dictates the charge and current distributions in the magnetosphere so that a gap has to adjust itself to produce the necessary current.  How could the polar cap maintain the global magnetospheric current is still not clear; the problem is widely discussed in the literature (see recent papers by \citet{arons07,beloborodov08,luo_melrose09,timokhin10}). In this paper, I suggest that the plasma flow breaks into separate narrow streams as a result of the electric current adjustment in the open field line tube.
%; in the opposite case, the magnetosphere gives back "extra" charges modifying the structure of the gap.

\section{Electric current adjustment in pulsar magnetospheres}
Within the magnetosphere, the plasma shields the longitudinal (along the magnetic field) component of the rotationally induced electric field providing at any point the charge density \citep{goldreich_julian69},
 \eqb
\rho_{GJ}=-\frac{\mathbf{\Omega\cdot B}}{2\pi c}.
 \label{GJ}\eqe
This condition is a generalization of the standard condition of the quasi-neutrality because in the corotating frame, the electric field vanishes when the charge density is equal to that of the Goldreich-Julian. Any deviation of the local charge density from (\ref{GJ}) results in a longitudinal electric field, which redistributes the charges to establish $\rho=\rho_{GJ}$ . %The necessary field is weak in the sense that the corresponding potential is small as compared with the total rotationally induced potential; this is because the energy of the secondary plasma particles is small as compared with the total potential.
Therefore in the plasma filled magnetosphere, the deviation of the charge density from that of Eq. (\ref{GJ}) remains negligibly small at any point;  one can say that the plasma shields the electric field in the corotating frame.

The electric current within the open field line tube is also not a free parameter. The current distribution is dictated by the global structure of the magnetosphere.
In the magnetosphere, the magnetic field lines bend backwards with respect to the rotation direction in order to allow the plasma to stream along them at a velocity smaller than the speed of light even though the rotation velocity becomes superluminal beyond the light cylinder. The current should be distributed such that the necessary magnetic field is maintained. Formally, the current along any magnetic field line is determined by the condition of the smooth transition of the flow along this field line through the light cylinder \citep{ingraham73, contopoulos99}. Within the plasma flow, the necessary current density is maintained owing to a small difference in the average electron and positron velocities. The current flows along the magnetic field lines and %the continuity condition requires that
the same current must flow across the polar gap at the base of the open field line tube. However, the flow in the gap is charge separated so that there is no additional degrees of freedom to adjust the current. The gap is filled only with the relativistically moving primary particles with the density close to $\rho_{GJ}/e$.
Therefore the polar gap models predict that the current density across the gap is close to $\rho_{GJ}c$.   Even though the current density in the self-consistent magnetosphere is of the order of $\rho_{GJ}c$ , it is not equal to this value; moreover the current density is not the same over the polar cap. For example in the axisymmetric magnetosphere, the current density is equal to $\rho_{GJ}c$ at the axis of the pulsar, decreases towards the periphery of the polar cap, changes sign near the boundary and again becomes zero at the boundary \citep{contopoulos99,timokhin06}. In the oblique case, the current has opposite signs in the upper and lower part of the polar cap (e.g., \citet{bai_spitkovsky10}).

The mismatch between the current in the gap and the current flowing along the same magnetic field line in the magnetosphere means that "extra" charges are accumulated somewhere above the polar gap until the longitudinal electric field arises sufficient to drive back extra charges. Since the particle energy in the flow is small as compared with the total rotationally induced potential, the necessary excess of the charge density over the Goldreich-Julian density is very small. Thus, one should expect that a reverse particle flow is easily formed \citep{lyub_adjust92}; simulations of the plasma flow within the open field line tube confirm this conjecture \citep{lyub_adjust09}. The reverse particle flow fills the gap from above and shorts out the longitudinal component of the electric field provided the sign of the particle charges is the same as the Goldreich-Julian charge. In nearly aligned magnetospheres, this occurs at the periphery of the polar cap where the magnetospheric current density is less than $\rho_{GJ}c$ or even has the opposite direction \citep{contopoulos99,timokhin06}. Therefore when the reverse flow reaches the surface of the star, the  particle acceleration is terminated and the pair production ceases.

This could result in a non-steady regime such that the required current is maintained only on average. Recently one-dimensional models were developed, both analytical \citep{levinson_etal05,beloborodov08,luo_melrose09} and numerical \citep{timokhin10}, for such a regime. These models assume that the gap height, $h$, is less than the polar cap radius and that the magnetospheric current is already fixed just above the gap. In this case one can imagine that different parts of the gap oscillate independently thus providing the necessary current distribution over the polar cap. If the gap height is larger than the polar cap radius, as it should be if the surface freely supplies the charges (e.g. \citet{hibsch_arons01b,hard_muslimov02,hard_muslimov_zhang02}), the situation should be much more complicated. Moreover, the formation time of the reverse flows, $\sim\Omega^{-1}$,  is much larger than the characteristic gap time and is generally different at different magnetic field lines. In this case, one can hardly imagine that the necessary current distribution over the polar cap is achieved via the gap oscillations.
%Another possibility is that in the three-dimensional case, an additional degree of freedom emerges, which could provide the current adjustment.

Another possibility arises if one abandons the one-dimensional picture at all. Namely, the cascade could become highly inhomogeneous across the polar cap such that the required current distribution is achieved only by averaging over the small scale structure.
One can speculate that the primary beam breaks into many subbeams separated by streams of the reverse plasma flow. The primary subbeams are accelerated in the electric gaps and produce electron-positron plasma streams; the reverse flows are formed in the outer magnetosphere via the charge and current adjustment. When a narrow reverse flow reaches the surface, the electric field is shielded at these field lines therefore the gap shifts so that the plasma is now produced at neighboring field lines. This implies a slow drift of the entire system of the upward and downward moving plasma streams (Fig.1). Recall that in the axisymmetric magnetosphere, the current density is equal to $\rho_{GJ}c$ at the axis \citep{contopoulos99,timokhin06} so that the reverse plasma flows arise only at the periphery of the polar tube. Therefore in pulsars with small inclination angles, the flow breaks into a steady central stream surrounded by upward and downward streams slowly circulating around the magnetic axis.  The primary subbeams carry the current density $\rho_{GJ}c$ and so do the plasma streams produced above the gaps. The reverse plasma flows could carry any current of any sign provided the pair density within them is at least a few times larger than the Goldreich-Julian density. If these streams are narrow enough, such a structure could provide any current distribution averaged over the scale larger than the stream width but smaller than the width of the open field line tube.

\begin{figure*}
\includegraphics[width=20 cm,scale=1]{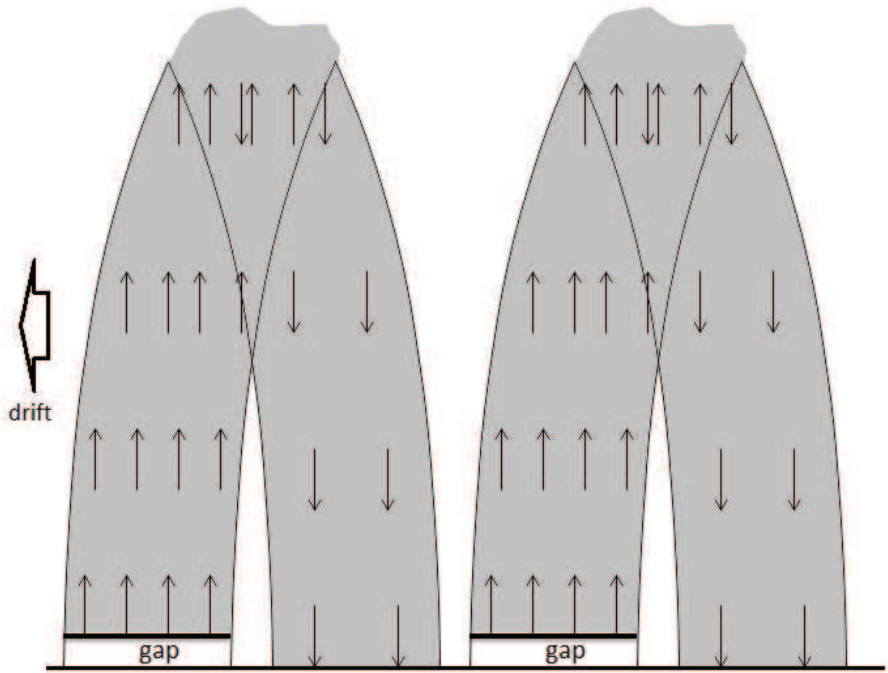}
\caption{A view of the plasma flows in the pulsar magnetosphere in the frame of the star. The flow breaks into a system of narrow streams. Each outward stream is produced in a corresponding potential gap where the primary particles are accelerated and give rise to the pair cascade. When the stream reaches the outer magnetosphere, a fraction of particles is redirected back because of the electric current mismatch between the polar gap and the magnetosphere. When the reverse plasma flow reaches the surface of the star, the electric field is shielded therefore the gap has to shift, which results in the  drift of the whole pattern (to the left in the picture). All the particles move along the magnetic field lines (strictly vertical in the picture) however, the plasma streams are bent because the polar gaps (particle sources) drift so that the plasma produced at some bundle of the magnetic field lines reaches large altitudes when the plasma producing cascade is already shifted to the neighboring magnetic lines.}
\end{figure*}

Even though this structure resembles that proposed by \citet{wright03}, the physics is different. Wright assumed that the potential drop in the gaps remains the same as in the Ruderman-Sutherland model. However, the reverse plasma flows shield the electric field therefore the potential drop in the gaps should be much smaller. In this case, the $\mathbf{E\times B}$ drift rate is very low and could be neglected; the pattern circulates because the gaps could not remain at rest: they have to shift before the corresponding reverse flows come to the surface. Therefore the subpulse drift is determined not by the $\mathbf{E\times B}$ drift rate but by the time necessary for the reverse flow to form and to reach the surface. The direction of the subpulse drift may be arbitrary.

An important point is that a low potential drop in the gap does not generally preclude the pair formation. It was realized some time ago \citep{kundt_schaaf93,sturner_etal95,luo96,zhang_qiao96} that the inverse Compton scattering of the thermal radiation from the star surface provides pair-producing gamma-rays at the energy of the primary electrons much smaller than a few TeV required by the classical models. Depending on the structure of the magnetic field, the energy 1-10 GeV is sufficient in order to produce, in the Klein-Nishina regime, gamma photons, which are easily converted into pairs \citep{hibsch_arons01a,hibsch_arons01b,hard_muslimov02,hard_muslimov_zhang02,arendt_eilek02,medin_lai10}.  In this case, the pair yield is significantly smaller than in the classical models however, the plasma density larger than the Goldreich-Julian density (multiplicity larger than unity) could be achieved at typical pulsar parameters if the magnetic field near the surface is not a dipole \citep{hibsch_arons01b}.

\section{Structure of the multi-beam polar cap accelerator}

In the picture outlined above, the polar cap region is organized into an assembly of narrow gaps producing the outwards plasma streams. The gaps are separated by reverse plasma flows. Since the reverse flows shield the electric field, the potential drop in the gaps is small, the narrower the gap the smaller the potential drop. The condition that the gaps produce the electron-positron plasma places a lower limit on the gap size. In this section I demonstrate that the radius of the gaps could be as small as 10\% of the polar cap radius; then the number of steams is compatible with the observed picture of drifting subpulses.

Even though a large potential drop is not necessary for pair production, the acceleration process could occur only if the electric force exceeds the radiative braking force \citep{kardashev_etal84,xia_etal85,daugherty_harding89,chang95,sturner95}. The strongest is the force due to the resonance cyclotron scattering.  Let the surface of the star radiate as a black body with the temperature $T$ and the particle move perpendicularly to the surface with the Lorentz factor $\gamma$. Then the resonance braking force is found as \citep{dermer90}
 \eqb
F=\frac{r_e\omega_B^2kT}{c^2\gamma}\ln\left[1-\exp\left(-\frac{\hbar\omega_B}{\gamma kT}\right)\right]^{-1};
 \eqe
where $\omega_B$ is the cyclotron frequency, $r_e$ the classical electron radius. The maximal force is
\eqb
F_{\rm max}=0.48\frac{r_e\omega_Bk^2T^2}{\hbar c^2}=0.057B_{12}T_6^2m_ec^2\,\rm cm^{-1};
\label{brake}\eqe
where $B=10^{12}B_{12}$ G is the surface magnetic field, $T_6=T/10^6\,{\rm K}$.
It is achieved at $\gamma=0.7\hbar\omega_B/kT=100B_{12}T_6^{-1}$ when the cyclotron frequency falls into the Lorentz shifted thermal peak. The primary particles could acquire larger energies if the electric field in the gap exceeds $F_{\rm max}/e$.

When the primary particles are extracted from the surface of the star, they are accelerated at first as in a plane-parallel slab. The electric field in this zone arises due to the space charge limited effect; it is estimated as \citep{michel74}
 \eqb
eE=\sqrt{8\pi e\rho_{\rm GJ}m_ec^2}=0.3B_{12}^{1/2}P^{-1/2}\,m_ec^2\,\rm cm^{-1}.
 \eqe
One sees that at typical parameters, the electric force exceeds the braking force (\ref{brake}).
This field is extended up to the altitude comparable with the radius of the primary beam, $a$, beyond which it decreases exponentially. In classical models,  the radius of the primary beam is equal to the polar cap radius. In our model, the primary beam is split into narrow subbeams separated by reverse plasma flows therefore in our case, $a$ is the radius of a subbeam. The field of the space charge accelerates particles up to the Lorentz factor $\gamma\sim eEa/m_ec^2=300B_{12}^{1/2}P^{-1/2}(a/10^3\,\rm cm)$.

Above this zone, the particle is accelerated because, due to the general relativistic frame dragging, a mismatch arises between the charge density of the primary beam and the local Goldreich-Julian charge density \citep{muslimov_tsygan92,muslimov_harding97},
 \eqb
\frac{\Delta\rho}{\rho_{\rm GJ}}=\frac{2GI}{c^2(R_*+z)^3}\approx\frac{0.5I_{45}}{R_{*6}^3}\frac z{R_*}.
 \eqe
Here $R_*=10^6R_{*6}$ cm is the radius of the star, $I=10^{45}I_{45}$ g$\cdot$cm$^2$ the moment of inertia of the star, $z$ the altitude above the surface (I assume that $z\ll R_*$). Within a narrow primary subbeam surrounded by conducting reverse plasma flows, the electric field is nearly transverse so that the electric potential is estimated as  $\phi=2\pi\Delta\rho a^2$. Then the accelerating electric field is
 \eqb
E=\frac{d\phi}{dz}=\frac{6\pi GI}{c^2R_{*}^4}a^2\rho_{\rm GJ}.
 \eqe
Numerically the accelerating force could be presented as
 \eqb
eE=6\frac{B_{12}}{P^2}\left(\frac a{a_0}\right)^2\,m_ec^2\,\rm cm^{-1}.
 \eqe
 Here
 \eqb
a_0=\sqrt{\Omega R_*^3/c}
 \eqe
is the polar cap radius found for the vacuum dipole field. Note that simulations of the pulsar magnetospheres \citep{contopoulos99,timokhin06,bai_spitkovsky10} show that the real radii of the polar caps are approximately 10\% larger than $a_0$.
 One sees that the accelerating force exceeds the braking force (\ref{brake}) if $a>0.1PT_6a_0$. Then the particles are accelerated up to large Lorentz factors sufficient for the pair production.

Due to the frame dragging effect, the primary particles could acquire Lorentz factors
 \eqb
\gamma\sim \frac{eER_*}{m_ec}=6\cdot 10^4 \frac{B_{12}}{P^2}\left(\frac a{0.1a_0}\right)^2.
 \eqe
For small $a$, these electrons could not produce pairs via curvature radiation however, the inverse Compton scattering of the thermal radiation from the surface of the star could initiate the pair cascade. The one-photon pair production occurs at $\epsilon\gtrsim 1000m_ec^2$. The resonance scattering provides photons with the energy $\epsilon=\gamma\hbar\omega_B=200\gamma_4B_{12}m_ec^2$. At $\gamma\gtrsim 10^4$, the non-resonant scattering in the Klein-Nishina regime produces more convertible photons than the resonance scattering does\citep{hibsch_arons01a}. Namely, the non-resonance braking force is \citep{blumenthal_gould70}
 \begin{eqnarray}
 F^{\rm NR}=\frac{\pi^2}3\left(\frac{kT}{m_ec^2}\right)^2\frac{\alpha^2}{\lambda_C}m_ec^2
\left(\ln\frac{4\gamma kT}{m_ec^2}-2\right) \nonumber\\
= 0.02T^2_6\left(\ln\frac{\gamma T_6}{1500}-2\right)m_ec^2\,\rm cm^{-1};
 \end{eqnarray}
where $\alpha$ is the fine structure constant, $\lambda_C=h/m_ec$ the Compton wavelength. In this regime, the energy of the scattered photons is close to the electron energy therefore when the electron with the Lorentz factor $\gamma\sim$ few$\times 10^4$ passes the distance $\sim R_*$, it produces $\sim F^{\rm NR}R_*/\gamma m_ec^2\sim 1$ photon, which in turn gives rise to a few pairs with a smaller energy (for more details, see \citet{hibsch_arons01a,hibsch_arons01b,hard_muslimov_zhang02}). One sees that even if the width of the primary subbeams is roughly ten times less than the width of the polar cap, the primary electrons produce enough pairs to maintain the charge and current densities required by the pulsar magnetosphere.

\section{Radiation pattern and drift rate.}
Now let us come round to the overall picture. As a result of the electric current mismatch between the polar cap accelerator and the magnetosphere, the flow in the open field line tube breaks into a system of the outward streams separated by the reverse plasma streams. The reverse streams reach the surface and short out the electric field forming plasma columns connecting the star with the outer magnetosphere. The electric gaps are formed between the reverse flows; the primary subbeams are accelerated in the gaps and produce the outward streams of the secondary plasma. The current density in the outward streams remains close to the Goldreich-Julian value, as is dictated by the properties of the polar cap accelerator. The "extra" current flows in the downward streams. The narrower the streams, the closer the current distribution to that demanded by the magnetosphere therefore one can assume that the polar cap is populated as densely as possible with the streams. According to the above estimates, the radius of the subbeam could be as small as $0.1a_0$. The radius of the reverse flows should be approximately the same therefore the distance between the centers of adjacent subbeams could be taken as $0.4 a_0$.

In this model, the potential drop is small therefore the $\mathbf{E\times B}$ drift could be neglected. The subbeams drift because they could not remain at rest as the electric field is shielded when the reverse plasma flow reaches the surface of the star. %Therefore the gap, and therefore the outward stream produced by the primary beam accelerated in the gap, has to shift.
An important point is that in aligned magnetospheres, the current is equal to $\rho_{GJ}c$ at the axis \citep{timokhin06} therefore in pulsars with small inclination angles, the reverse flow is not formed near the axis of the polar cap so that the central subbeam could remain at rest. In this case other subbeams could circulate around the center of the polar cap. Taking into account the above estimate of the minimal distance between the subbeams, one can place the streams on the polar cap such that they form up to two concentric rings and a stream in the center of the cap. This is compatible with the basic observed radiation pattern: an axial (core) beam surrounded by two nested cone beams. The number of streams in the outer ring could be estimated as $2\pi a_0/(0.4a_0)=16$, which is compatible with the number of subbeams (20) found observationally by \citet{deshpande_rankin01,mitra_rankin08}.

The circulation velocity is determined by the time necessary for the plasma stream reach the outer magnetosphere, where the electric current mismatch produces the reverse flow, and then by the time necessary for the reverse flow to reach the surface of the star.
Let us estimate the time necessary to reach the outer magnetosphere and come back. The particles move along the rotating magnetic field lines so that their velocity could be presented as
 \eqb
\mathbf{v}=\mathbf{\Omega\times r}+\xi\mathbf{B}.
 \eqe
The constant $\xi$ is found from the condition that the velocity is close to the speed of light, which yields
 \eqb
\xi_{\pm}=\frac{-B_{\phi}x\pm\sqrt{B^2-B_p^2x^2}}{B^2};
 \eqe
where $x=\Omega r/c=2\pi r/cP$ is the cylindrical radius in units of the light cylinder radius, $P$ the pulsar period, $B_p$ and $B_{\phi}$ the poloidal and azimuthal components of the magnetic field, respectively. The sign $\pm$ refers to outward and downward motion, respectively. Note that if $\Omega$ and $B_p$ are positive, $B_{\phi}$ is negative. The time necessary to reach a point at the cylindrical distance $x_1$ from the axis or to come back from this point to the star is
 \eqb
t_{\pm}(x_1)=\pm\int_0^{x_1}\frac{dx}{\xi_{\pm}B_x};
 \eqe
where the integral is calculated along the field line. The drift rate is determined by the condition that the gap shifts by its diameter, $2a$, for the time $t_++t_-$. It is customary to to characterize the drift rate by the time interval $P_2$ between two consecutive subpulses to pass the same longitude. In our model, $P_2=2(t_++t_-)$.

The reverse flows are formed due to the electric current mismatch between the magnetosphere and the polar cap. Since the magnetospheric current distribution is determined by the conditions of smooth transition of the flow through the light cylinder  \citep{ingraham73,contopoulos99} the reverse flows should be formed near the light cylinder. Note that $\xi_-$ goes to zero at the light cylinder so that $t_-(x_1)$ diverges as $x_1\to 1$ therefore the value of $P_2$ is determined by the details of the reverse flow formation. Just for the estimate, I calculated $t_{\pm}$ for the flow along the boundary between the closed and open parts of the magnetosphere in the  axysimmetric model by \citet{timokhin06}. The numerical data were kindly presented by Andrey Timokhin. I obtained $P_2=1.1P$ for $x_1=0.9$, $P_2=2P$ for $x_1=0.95$ and $P_2=3.8P$ for $x_1=0.97$. Therefore the typically observed $P_2\sim 2P$ could be easily obtained in the scope of this model. A significantly larger $P_2$ could be obtained only if the reverse flow is formed extremely close to the light cylinder.

\section{Conclusions}

In this paper, the subpulse modulation of the pulsar radio emission is attributed to the current mismatch between the magnetosphere and the polar cap accelerator.
The current distribution in the magnetosphere is determined by the global electrodynamics. For example in the axisymmetric magnetosphere, the current density is equal to $\rho_{GJ}c$ at the axis and decreases towards the periphery of the open field line tube\citep{contopoulos99,timokhin06}. The problem is that the polar cap accelerator could inject only the current close to $\rho_{GJ}c$. % Because of a very large inductance of the magnetosphere, any extra charges should be expelled by the Lenz's electric field.
The "extra" charges should be expelled from the magnetosphere, which implies the formation of reverse particle flows \citep{lyub_adjust09}.
%which fall from above onto the polar gap.
If the "demanded" current is less than the injected one, the reverse flow brings back charges of the same sign as $\rho_{GJ}$. These charges fill the polar gap and shield the longitudinal electric field there thus arresting the plasma production. This could in principle lead to a spasmodic plasma production \citep{levinson_etal05,luo_melrose09,timokhin10}. % however, one could hardly imagine such a regime when the characteristic time of the reverse flow formation is different at different field lines and in any case significantly exceeds the characteristic gap time-scale.
Here I suggest that the adjustment of the electric current between the magnetosphere and the polar cap accelerator occurs via the formation of narrow streams transferring different currents.

In this scenario, the plasma is generated in narrow gaps scattered over the polar cap, which yields a system of narrow upward moving streams with the current density close to $\rho_{GJ}c$. In the light cylinder zone, a fraction of this plasma is redirected back forming reverse plasma streams. When the reverse streams reach the surface of the star, the plasma production ceases at the corresponding bundles of magnetic field lines therefore the plasma producing gaps have to shift, which results in the slow circulation of the pattern. The current in the reverse streams is not restricted therefore the overall current distribution averaged over the narrow streams could easily match that demanded by the magnetosphere. In this model, the subpulse modulation arises naturally. The transverse size of the streams is limited from below by the condition that the plasma could be produced in the gaps. I demonstrated that the plasma production process is still possible if the radius of the streams is as small as 10\% of the polar cap radius. In this case, 10 - 20 subbeams could be placed along the circumference of the polar cap, which is compatible with the observations of drifting subpulses \citep{deshpande_rankin01,mitra_rankin08}. Taking into account the estimated size of the streams, one sees that they could be arranged into a system of two concentric rings and a central stream, thus reproducing a generic radiation pattern: the core beam surrounded by two nested cones \citep{rankin83,rankin93}.

\acknowledgements
I am grateful to Andrey Timokhin for presenting me the results of his simulations of the pulsar magnetosphere. This work was supported by the Israeli Science Foundation under the grant 737/07.

%\bibliographystyle{spr-mp-nameyear-cnd}

%\bibliographystyle{apj}
%\bibliography{pulsar}

\hyphenation{Post-Script Sprin-ger}

\end{document}